\documentclass[preprint]{vgtc}               

\graphicspath{{figures/}{pictures/}{images/}{./}} 

\usepackage{times}                     

\usepackage{tabu}                      
\usepackage{booktabs}                  
\usepackage{lipsum}                    
\usepackage{mwe}                       

\usepackage{mathptmx}                  
\usepackage{tcolorbox}
\usepackage{xcolor}
\usepackage{listings}
\usepackage[table]{xcolor}
\usepackage{amsmath,amsfonts}
\usepackage{graphicx}  
\usepackage{array, multirow, booktabs}
\usepackage[nocompress, sort]{cite}

\newcolumntype{C}[1]{>{\centering\arraybackslash}p{#1}}

\definecolor{result_red}{HTML}{FFA9AD}
\definecolor{result_orange}{HTML}{FFD5A8}
\definecolor{result_green}{HTML}{9CD9B6}

\definecolor{lightred}{RGB}{255, 175, 177}
\definecolor{lightorange}{RGB}{255, 215, 174}
\definecolor{lightyellow}{RGB}{255, 250, 194}

\definecolor{kim_green}{HTML}{4CBB17}
\definecolor{yalong_purple}{HTML}{9300FF}
\definecolor{ari_red}{HTML}{EC1313}

\definecolor{pending_user_study_res}{HTML}{2171DF}

\newcommand{\para}[1]{\vspace{0.35em}\noindent\normalsize\textbf{#1.}\xspace}

\newcommand{\sysmemento}[0]{Memento\xspace}
\newcommand{\rsam}[0]{RSAM\xspace}
\newcommand{\gpt}[0]{GPT-4o\xspace}
\newcommand{\gemini}[0]{Gemini 2.0-Flash\xspace}

\onlineid{5576}

\vgtccategory{System}

\vgtcinsertpkg

\usepackage{orcidlink}

\preprinttext{%
  \ifodd\value{page}%
    \parbox{0.9\textwidth}{%
      © 2026 IEEE. This is the author's version of the article that will appear at the IEEE Conference on Virtual Reality and 3D User Interfaces Abstracts and Workshops (IEEE VRW). The final version of this record is available at: \href{https://doi.org/10.1109/VRW70859.2026.00180}{\textcolor{blue}{\textit{10.1109/VRW70859.2026.00180}}}
    }%
  \fi
}

\title{Memento: Towards Proactive Visualization of Everyday Memories with Personal Wearable AR Assistant}

\author{
Yoonsang Kim\thanks{e-mail:yoonsakim@cs.stonybrook.edu}\\
\scriptsize Stony Brook University
\and
Yalong Yang\thanks{e-mail:yalong.yang@gatech.edu}\\
\scriptsize Georgia Institute of Technology
\and
Arie E. Kaufman\thanks{e-mail:ari@cs.stonybrook.edu}\\
\scriptsize Stony Brook University
}

\teaser{
  \centering
  \vspace{-2mm}
  \includegraphics[width=\linewidth]{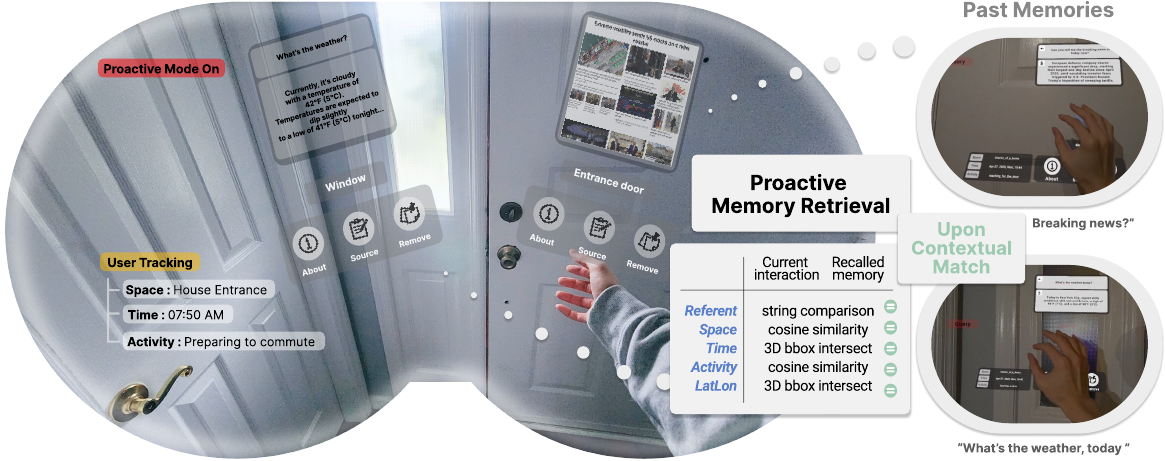}
  \vspace{-6mm}
  \caption{Concept illustration of \sysmemento. The user's prior queries are stored in the form of ``memories'', along with their referent, spatiotemporal, and activity contexts. These memories are proactively situated onto daily environments upon alignment with the user's current interaction context, serving as a visual reminder and for learning.}
  \vspace{-1.5mm}
  \label{fig:teaser}
}

\abstract{
We introduce \sysmemento, a conversational AR assistant that permanently captures and memorizes users' verbal queries alongside their spatiotemporal and activity contexts. By storing these ``memories,'' \sysmemento discovers connections between users' recurring interests and the contexts that trigger them. Upon detection of similar or identical spatiotemporal activity, \sysmemento proactively recalls user interests and delivers up-to-date responses through AR, seamlessly integrating the AR experience into their daily routine. Unlike prior work, each interaction in \sysmemento is not a transient event, but part of a connected series of interactions with coherent long-term perspective, tailored to the user's broader multimodal (visual, spatial, temporal, and embodied) context. We conduct a preliminary evaluation through user feedback from participants of diverse expertise in immersive apps, and explore the value of a proactive context-aware AR assistant in everyday settings. We share our findings and challenges in designing a proactive, context-aware AR system.
}


\begin{document}


\firstsection{Introduction}

\maketitle
Recent advances in Large Language Models (LLMs) have significantly transformed user interfaces, enabling natural-language conversations that capture the semantics and nuances of user queries and generating rich, context-aware responses. Coupled with AR (Augmented Reality), LLM-powered conversational assistants not only deliver dynamic, real-time answers, but also ground digital content within the physical environment, creating interactions that empower intuitive engagement with real-world contexts.

Existing systems~\cite{chang2024worldscribe, dogan2024augmented, lee2024gazepointar} have demonstrated the potential of combining AR, conversational interface, multi-sensory data, and LLMs to enable immersive, context-aware interactions. These approaches enrich user experiences by overlaying digital content onto everyday environments. However, these systems primarily focus on responding to immediate queries based on current or recent visual inputs, and short-term conversational history. While suitable for generic verbal queries, they fall short of long-term, personalized contexts. Also, the response to a question about an object can vary significantly since the meaning of the object can carry different value across users, and can shift according to spatial (\textit{Where}), temporal (\textit{When}), and task-driven (\textit{What}) contexts. For example, in existing systems, querying about a toy doll would output the cost and the property of the product, regardless of its spatiotemporal context, while it may possess different personal meaning at home--``\hyperref[fig:memento_design_concept]{my kid's favorite toy}''--compared to a store. Thus, a comprehensive understanding of the context behind a user's query is necessary. Another set of research explores personal ``contextual memory banks'' 
that assist users in recalling prior embodied interactions when queried~\cite{li2024omniquery, omniactions, shenencode24}. Yet, there remain opportunities to expand the use cases beyond user-initiated (reactive) queries, towards proactive, and immersive, situated visualization of memories.

Moreover, we focus on the growing research interest in everyday smart wearable technologies (e.g., HMDs, glasses, earbuds, watches) that lifelog users' actions and surroundings~\cite{arakawa2024prism, fang2025mirai, vinci, shenencode24, yang2025egolife, zulfikar2024memoro}. These works hint at how future personal device form factors can capture patterns and contexts of users' daily activities, and memorize them for remembrance or insight assistance.

To this end, we introduce \textit{\textbf{\sysmemento}}, a conversational AR assistant that aims to explore the usability and potential of proactive visualizations combined with contextual memories on a wearable AR form factor. \sysmemento captures and memorizes a user's verbal queries (\cref{fig:teaser}--``What is the weather like today'', ``Tell me the breaking news of the day''), along with their spatiotemporal and activity contexts (\cref{fig:teaser}--``House entrance'', ``at 7:50AM'', while ``Preparing to commute''). Utilizing this personal ``contextual memory,'' \sysmemento identifies the user's daily routine and proactively proposes visualizations of the user's interests learned from prior memories. This enables the users to bypass an explicit query amid repeated routines and serves as an informative visual reminder, as well as a new learning opportunity within everyday environments. 

We pose `\textit{\textbf{proactivity}}' as one potential interaction pattern of everyday wearable form factors--beyond user-requested unidirectional information flow--given their capability to memorize recurring behaviors and predict users' needs through multimodal AI-driven assistance~\cite{fang2025mirai, liu2024proactive, zulfikar2024memoro}. By designing a prototype proactive context-aware system, \sysmemento, and deploying it in various real-world scenarios, we share preliminary insights of everyday use of wearable AR. Our main contributions are:
\begin{itemize}
\vspace{-2mm}
\item We design a \textbf{proactive AR assistant} prototype that situates visualizations of user's prior interests (e.g., weather, traffic) without explicit requests, serving as a smart visual reminder and live information provider in daily environments.

\vspace{-1.7mm}
\item We introduce the concept of \textbf{Referent-anchored Spatiotemporal Activity Memory (RSAM)} representation, which anchors user interactions to everyday objects along with their spatial, temporal, and activity context. RSAM enables \sysmemento to surface long-term personal interests and recurring intentions, supported by a lightweight retrieval mechanism for recalling contextually relevant memories.

\vspace{-1.7mm}
\item We share \textbf{insights on the applicability of proactive, context-aware AR assistant} in real-world scenarios via a simplified prototype and report insights into designing such a system.

\vspace{-2mm}
\end{itemize}

\section{Related Work}

\subsection{AI and AR Assistance in Routine Daily Life}
Recent systems such as MaRginalia~\cite{qiu2025marginalia}, XR-Objects~\cite{dogan2024augmented}, and Mirai~\cite{fang2025mirai} showed that linking digital content to physical referents and past interactions helps users continue tasks without re-finding information. With emerging wearable XR devices, XR is increasingly positioned as an everyday interface that maintains continuity between physical environments and digital memory~\cite{lv2024aria, azuma2019road, barfield2015fundamentals, bressa2022data, tran2025wearable}. Memoro~\cite{zulfikar2024memoro} intelligently completes sentences and assists the user through audio using prior interactions.

The daily use of portable devices is highly routine~\cite{almourad2024exploring, dengmeasuring, kim2019real, sarker2019, silva2018discovering}, yet, current assistants require users to repeat similar queries daily and are reactively triggered. Proactive systems resurface information when contexts recur, enabling continuation of interactions without restarting. Satori~\cite{li2025satori} anticipates needs from task-based contexts, and Sensible Agent~\cite{lee2025sensible} proactively recognizes the needs of users for unobtrusive use of XR in public and minimizes interaction burden.

As XR transitions toward everyday use, proactively resurfacing personal context without repeated requests aligns with how users revisit routines, spaces, and objects. This motivates treating past interactions as personal ``memories'' linked to spatial, temporal, and activity cues, and resurfacing them whenever similar routines return, integrating digital recall into daily behavior.


\subsection{Context-aware AI Assistant with LLM}
LLMs have transformed user interfaces by enabling conversational, context-aware assistance. They are increasingly integrated into data analysis and visualization workflows~\cite{choe2024enhancing, text2viz, han2024deixis, shen2024data}, as well as immersive applications, where LLMs function as intelligent companions~\cite{de2024llmr, wang2025towards}. GazePointAR~\cite{lee2024gazepointar} demonstrates how combining gaze, gestures, and conversational history can disambiguate user queries through richer contextual understanding. Similarly, WorldScribe~\cite{chang2024worldscribe} addresses accessibility by generating textual scene descriptions for visually impaired users.

These works illustrate how LLMs enhance AR by grounding multimodal inputs into context-aware responses. \sysmemento builds on this by shifting from reactive assistance to proactive recall of prior user queries, enabled through memory lifelogging~\cite{remberingmemory, shenencode24} and context matching.

\subsection{Capture and Retrieval of Memories}
In situated visualization, the relationship between the user and the physical referents they interact with carries significant contextual value. In this work, we define each interaction--capturing spatial, temporal, verbal, visual, and activity context--as a ``memory''~\cite{barfield2015fundamentals, omar2025, liu2024investigating, satriadi2023proxsituated, satriadi2023context}. Effectively storing and retrieving these memories requires robust data management. Traditional techniques such as R-trees and Oct/Quad-trees have been adopted to handle large-scale datasets~\cite{koutroumanis2021scalable, qi2020effectively, tian2022survey}, and are efficient for geo-tagged data, but struggle with unstructured queries or those requiring semantic context fusion~\cite{liu2024suql, caesura}. Recent research leverages vector embeddings from LLMs to represent multimodal data, enabling flexible retrieval through natural language and allowing systems to synthesize semantically coherent responses~\cite{asai2023retrieval, chen2022murag, kashmira2024graph, khan2024situational, lewis2020retrieval}. 

We build on this by integrating spatiotemporal indexing with embedding-based search, supporting real-time memory retrieval that balances structural filtering and semantic understanding.

\subsection{User Activity and Context Modeling}
Modeling user activity and context is essential for delivering intelligent and timely AR assistance. Prior research leverages multimodal data, including gaze, gesture, speech, and spatial context, to infer user intent~\cite{fang2025mirai, vinci, liu2024proactive}. Kim et al.~\cite{kim2025explainable} demonstrate that combining user actions, dialogue, and spatiotemporal information enables LLMs to extrapolate intent. Memoro~\cite{zulfikar2024memoro} introduces query-less assistance by detecting conversation topics and augmenting users' remembrance through audio feedback. OmniActions~\cite{omniactions} utilizes multi-sensory inputs to anticipate user actions.

\begin{figure}[!t]
    \centering
    \includegraphics[width=\linewidth]{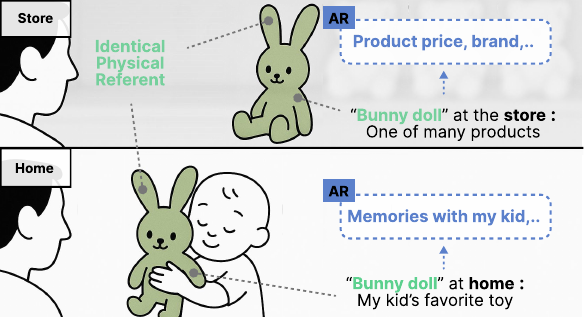}
    \vspace{-6.5mm}
        \caption{The view of situated AR in \sysmemento. A physical referent derives its meaning from its spatiotemporal context (\textit{Where} and \textit{When} of the object). AR content anchored to it serves as an augmentation layer that extends the physically visible information of the referent, within the same context. The referent acts as the bridge between the physical and virtual worlds.}
    \vspace{-2mm}
    \label{fig:memento_design_concept}
\vspace{-4mm}
\end{figure}

Building on these insights, we capture user activities with their personal context to leverage them as a ``personal spatial memory''. The memories act as triggers for surfacing information when the context is semantically aligned--each spatiotemporal-activity-referent context holds disparate personal memories (\cref{fig:memento_design_concept}).


\section{\sysmemento: Proactive Memory-aware Assistant}

\begin{figure*}[!t]
    \centering
    \includegraphics[width=\linewidth]{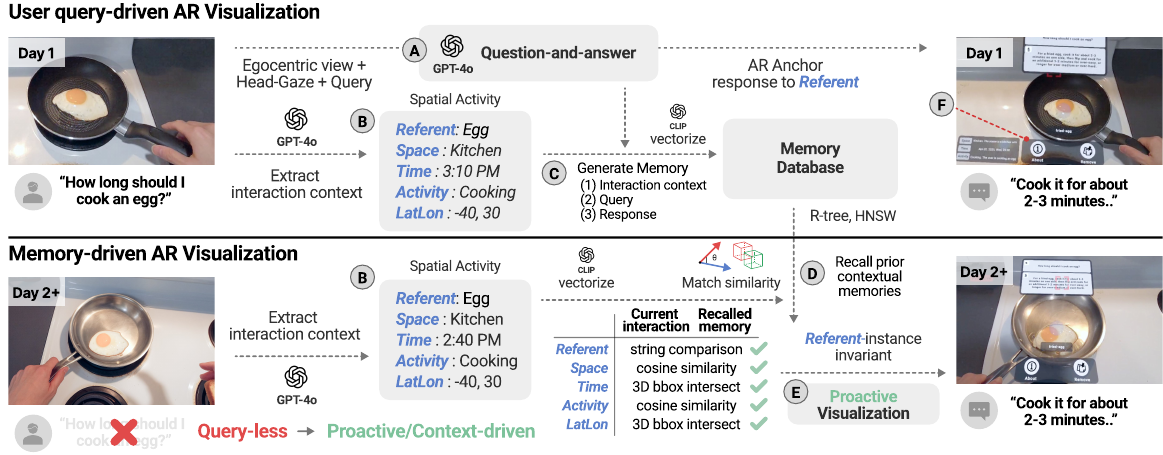}
    \vspace{-8mm}
    \caption{
    \sysmemento pipeline overview: With a verbal query, the visual, head-gaze, and the transcribed textual query from the user are sent to \sysmemento. Upon completion of response generation, the answer is situated onto the referent that the query is initiated to, and is permanently stored along with its spatial activity context as a form of a memory. After the initial query, the memory is proactively recalled when the user's similar contextual spatial activity, and situated onto the referent of the user's prior interest, without an explicit query from the user. 
    \includegraphics[height=1em]{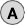}~\hyperref[subsubsec:personal_assistant]{Personal Assistant}, 
    \includegraphics[height=1em]{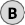}~\hyperref[subsubsec:spatial_activity_recognizer]{Spatial Activity Recognizer}, 
    \includegraphics[height=1em]{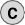}~\hyperref[subsubsec:memory_generation]{Memory Generator}, 
    \includegraphics[height=1em]{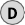}~\hyperref[subsubsec:memory_retrieval]{Memory Retriever}, 
    \includegraphics[height=1em]{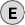}~\hyperref[subsubsec:proactive_visualization]{Proactive Visualizer}, 
    \includegraphics[height=1em]{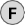}~\hyperref[subsubsec:adjustment_inferface]{Adjustment Interface}}
    \vspace{-1.25mm}
    \label{fig:pipeline}
\vspace{-4mm}
\end{figure*}

\vspace{-1mm}
\subsection{Design Considerations}
\label{sec:design_considerations}
Designing an AR system for everyday use requires principles that support convenience, seamlessness, and robustness across routine contexts. Prior work in ubiquitous and immersive computing highlights the need for systems that move beyond user-initiated interactions toward pervasive assistance~\cite{azuma2019road, grubert2016towards}. It shows that proactive support can reduce cognitive effort and improve continuity across repeated behaviors~\cite{andolina2018searchbot, liu2024proactive, meurisch2020exploring}. Wearable AR devices further enable embodied, egocentric interaction and situated visualization directly within the user's viewpoint~\cite{lee2023design, yang2025egolife}, while practically deploying them in-the-wild requires open-world scene understanding capable of identifying objects beyond predefined data class distribution~\cite{Cheng2024YOLOWorld,liu2024grounding,peng2023kosmos}. These insights motivate our design considerations:

\para{Proactive assistance}
To reduce repeated explicit queries for routine tasks~\cite{andolina2018searchbot,liu2024proactive}, we employ semi-proactive resurfacing. Assistance is triggered by recurring user behaviors while allowing dismissal to avoid information overload~\cite{liu2023visual, morris2023wearable, zulfikar2024memoro}. A context-aware memory bank stores interaction histories and resurfaces them under matched spatiotemporal and activity cues, with an adjustable interface providing users agency over repetition cycles~\cite{meurisch2020exploring}.

\para{Egocentric visualization}
Wearable AR serves as the primary platform to align digital content with users' perspectives and enable natural interaction (e.g., hand, gaze) in everyday environments. Compared to handheld AR, head-mounted displays more reliably capture multimodal egocentric inputs~\cite{yang2025egolife}, supporting situated visualization without restricting physical interaction.

\para{Open-world scene analysis}
Everyday AR requires recognizing objects users reference, even when they are unseen during training. To address this, \sysmemento leverages open-vocabulary object detection (YOLO-World~\cite{Cheng2024YOLOWorld}) and multimodal LLMs~\cite{gemini,gpt4v1} to identify objects from visual and textual descriptions, supporting assistance in unstructured, real-world environments.

\vspace{0.5mm}

While our design lays the groundwork for a full-fledged proactive assistant, in this work, we leverage our prototype system for simplified use case scenarios. \sysmemento acts as a context-driven reminder that resurfaces relevant information when routines repeat (beyond simple time-based notifications) serving as a step towards long-term proactive, context-aware AR support.

\vspace{-0.5mm}
\subsection{Prototype System Design}
\vspace{-0.25mm}
\para{Implementation Settings}
At the time of our prototype development, raw access to a camera sensor (with RGB color) with AR content display, was limited on glasses form factors (e.g., Meta Ray Ban's), while these form factors are more suitable for daily use of AR. Therefore, we focus on the exploratory use-cases of daily AR assistance rather than form factor comparisons, and deploy \sysmemento on a Meta Quest3 HMD (with Passthrough AR mode), and collect and discuss user feedback for preliminary insight building. \sysmemento is developed with Unity AR Foundation. The HMD is connected to a custom remote Python server, which offloads compute-heavy processes including LLM interactions. In addition, the HMD is network-paired with an auxiliary device (smartphone) to provide accurate GPS location of the user. The storing and recalling of memories are crucial internal operations of \sysmemento. We leverage two multimodal LLM models: \gpt~\cite{gpt4v1} and \gemini~\cite{gemini}, for generating memories, textual descriptions of scenes/activities, and referent localization -- and CLIP embedding model~\cite{radford2021learning} to vectorize memories, and deterministically compare retrieved memories.

\vspace{-1mm}
\subsubsection{Personal Conversational Assistant}
\label{subsubsec:personal_assistant}
The Assistant module supports the other five modules of \sysmemento. As shown in \cref{fig:pipeline}, the assistant acts as the starting point that enables a user to interact with a referent in the physical environment. We track the user's head-gaze and pinch gesture to identify the user's referent of interest. At any point during an AR session, the user can perform the head-gaze-and-pinch to inquire about a topic from the assistant and receive an answer. Upon the detection of a pinch gesture, we capture the user's view as well as a verbal query, which is transcribed into text with the Whisper model~\cite{whisperUnity}, and transmitted to the LLM, connected to our server. We also place a visual marker on the 2D view capture to provide visual guidance to the referent of the user's query~\cite{lee2024gazepointar}. The marker is generated by projecting the 3D position of the referent derived from the gaze, onto the 2D capture. Moreover, to respond to the user's queries with live information (beyond the static knowledge-base of LLMs), we integrate the Google Custom Search API~\cite{googlecustomsearch} into the assistant.
\vspace{-1.1mm}

\subsubsection{Spatial Activity Analysis}
\label{subsubsec:spatial_activity_recognizer}
Timely recognizing the user's current space, and identifying the activity the user is engaged in, is pivotal to \sysmemento. The appropriate inference of this information enables proactive visualizations of context-relevant AR content. We lifelog the user's first-person view (via the external camera of the HMD), latitude, longitude, and timestamp, in a continuous manner, and infer the user's current scene and activity. This spatial activity analysis maintains an up-to-date context of the user's behaviors in the real-world. The interaction contexts include, the user's immediate space, activity, time of activity, the referent of the activity, and the geographical location of the user. 

A textual description may lack sufficient details to fully characterize the user's spatial context. For example, two locations, both described as ``Kitchen'' may be in two geographically distinct locations (e.g., kitchens of different households). Thus, we first use \gpt to represent the user's current spatial activity in text, and associate geographical coordinates (latitude, longitude, accuracy: margin-of-error) with the Spatial Activity Profile. This combination of spatial activity contexts (latitude, longitude, visual-text descriptor, activity) enables \sysmemento to distinguish spaces of the real-world spaces, while preserving the fine-grained semantics of each space and activity. The descriptor which is a single vector representation, is generated by prompting \gpt to output textual description of the scene given an image capture of the user's view, and converted to CLIP-encoded embeddings.
\vspace{-1mm}

\begin{figure}[!t]
    \centering
    \includegraphics[width=\linewidth]{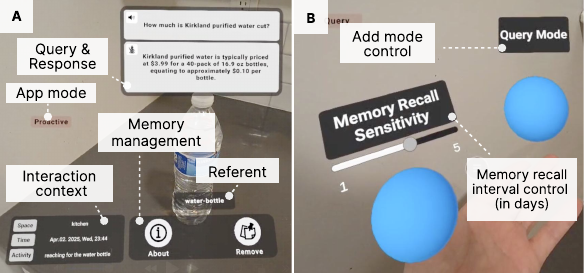}
    \vspace{-7mm}
    \caption{User interfaces of \sysmemento. (A) Example visualization of RSAM situated onto a `Water bottle' referent. (B) Context adjustment interface provides a switch between Query (New query) and Proactive (Prior query/Memory recall) mode, and adjust the interval of memory recalls (How often memory is recalled).}
    \label{fig:user_interface_explanation}
\vspace{-6mm}
\end{figure}

\subsubsection{Personal Memory Generation}
\vspace{-0.5mm}
\label{subsubsec:memory_generation}
The referent is at the heart of the memory that bridges AR visualization with the physical realm (\cref{fig:memento_design_concept}). Upon initial query interaction with a physical referent via \hyperref[subsubsec:personal_assistant]{Personal assistant}, the query and its metadata are recorded in the form of a memory. It holds the moment of the verbal query, the response, the source of the response, and the context of the spatial activity. We term this, \textbf{{R}}eferent-anchored \textbf{{S}}patiotemporal \textbf{{A}}ctivity \textbf{{M}}emory (\rsam).

To generate an \rsam, we first identify the referent in question, and retrieve the most recent \hyperref[subsubsec:spatial_activity_recognizer]{Spatial Activity Profile}. Then, we bring the user's verbal query, its response, and the response source together, and form a memory.
We store a new \rsam in a data structure that combines R-tree~\cite{rtree} with a graph-based hierarchical $K$-approximate nearest neighbor search algorithm known as HNSW~\cite{malkov2018efficient}. This hybrid structure enables fast insertion and retrieval of \rsam, suitable for an interactive AR assistance system. The R-tree provides coarse partitioning of data based on latitude, longitude, and timestamp keys, while fast retrieval of the vector-represented spatial activity context captures the semantics of the user's scene and activity via HNSW. 
\vspace{-0.5mm}

\subsubsection{Context-aware Memory Retrieval}
\vspace{-0.5mm}
\label{subsubsec:memory_retrieval}
Memory recall begins by accessing the most recent Spatial Activity Profile. It contains the user's current geographical location, scene information, activity, and timestamp. By filtering the existing {\rsam}s by geographic coordinates and time first, and hierarchically retrieving vectorized spatial activity, \sysmemento enables fast top-$K$ context similarity retrieval in a deterministic way. 

In an effort to reduce false positive memory suggestions upon the retrieval of {\rsam}s, we equally weigh the significance of each dimension in the Spatial Activity Profile. We sequentially verify the semantic alignment between each dimension, and the user's current context. The proximity of spatiotemporal context is assessed as the first step, then we verify the existence of the \rsam referents in the user's view, and finally fine-grained spatial activity semantics matching. Only when all dimensions are satisfied, an \rsam is suggested to the user.
\vspace{-0.45mm}

\begin{figure*}[!t]
    \centering
    \includegraphics[width=\linewidth]{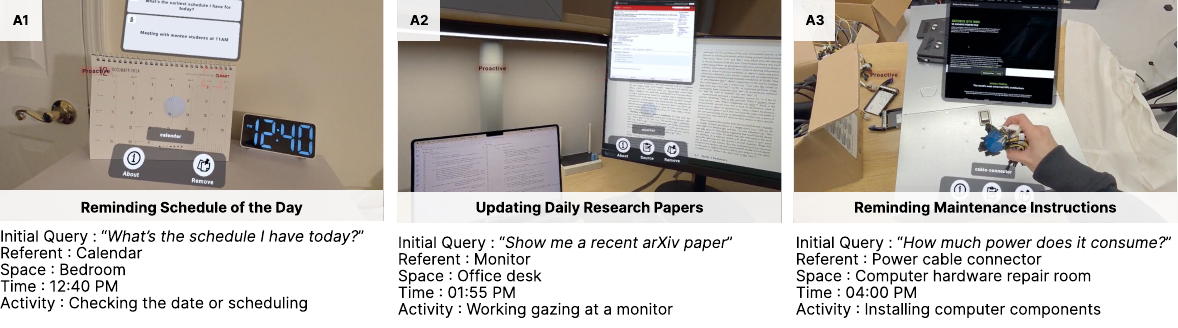}
    \vspace{-7.5mm}
    \caption{The three real-world use-cases of \sysmemento. Upon the contextual alignment of the user's current spatial activity (Referent, Space, Time, Activity), \sysmemento proactively recalls a prior RSAM and situates via AR.}
    \vspace{-2mm}
    \label{fig:usecase_and_applications}
\vspace{-3.5mm}
\end{figure*}

\subsubsection{Proactive Situated Information Display}
\label{subsubsec:proactive_visualization}
\vspace{-1mm}

\para{Referent detection}
Once the $K$-similar {\rsam}s--the past memories that align with the user's current spatiotemporal and activity context--return, the Proactive visualizer localizes the referents where the memories were situated previously. This situating of AR content is done by localizing the referents specified in {\rsam}s via an open-vocabulary detector. If present in the current view, the 2D center points are transmitted to the user's HMD to compute the 3D anchor points using the depth map acquired from the external camera of the HMD, at the time of the image capture.\vspace{-0.25mm}

\para{Content visualization}
A memory is AR-anchored onto the referent specified by an RSAM. As shown in \hyperref[fig:user_interface_explanation]{\cref{fig:user_interface_explanation}-A}, users are given the option to view the system reasoning on Spatial Activity contextual alignment. The Space, Time, and Activity of the matched information in a human-readable format is shown for explainability. The `Remove' option allows permanent dismissal of unwanted (e.g., mismatched) memories or one-time queries that do not require repeating. \sysmemento also proactively self-update the data of the contents providing newly available information on each repeat cycle. This makes \sysmemento, beyond a simple immersive reminder, a fresh information-fetching agent.
\vspace{-0.25mm}

\para{Content interaction}
To interact intuitively with the memories, we employ ``head-gaze-and-pinch'' and the design pattern of ``Head-gaze and dwell''~\cite{microsoftheadgazedwell}. Users can naturally look towards (head-gaze) the memory of interest, and select it with the pinch gesture. It not only provides an embodied way of interacting with data, but also serves as a two-step safeguard to confirm the desired memory of interest before initiating the raycast selection. 
\vspace{-0.25mm}

\subsubsection{User Preference and Proactivity Adjustment Interface}
\label{subsubsec:adjustment_inferface}
Fully autonomous information suggestion systems risk diminishing the user's sense of ownership and control~\cite{liu2023visual}. As discussed in~\cref{subsubsec:proactive_visualization}, \sysmemento presents a user interface that indicates the spatial activity of the user's situated memory (space, time, activity), and the option to remove the memory (\hyperref[fig:user_interface_explanation]{\cref{fig:user_interface_explanation}-A}). Furthermore, the level of proactivity of AR visualizations can vary depending on personal preferences and tasks~\cite{meurisch2020exploring}, thus, we provide an interface to adjust the memory recall sensitivity (\hyperref[fig:user_interface_explanation]{\cref{fig:user_interface_explanation}-B}) that controls the frequency of recall (How often the system recalls), in days.
\vspace{-1mm}

\section{Example Use Cases and Applications}
\vspace{-0.25mm}
\sysmemento leverages recurring patterns of daily activities to proactively surface relevant information without requiring repeated explicit queries. Rather than prompting the user to ask again, the system infers when past needs are likely to reappear based on recurring spatiotemporal routines and resurfaces the most contextually relevant memories through situated visualization on everyday objects. Below are three representative patterns of daily use, observed in our user study (denoted \hyperref[fig:usecase_and_applications]{A1} to \hyperref[fig:usecase_and_applications]{A3} across the sections and in \cref{fig:usecase_and_applications}).

\para{Managing daily routines through recurring checks (\hyperref[fig:usecase_and_applications]{A1})}
Users frequently request information tied to the start of their day, such as reviewing personal schedules or checking weather conditions before leaving home. \sysmemento captures such explicit queries once, and upon detecting similar activity patterns at a later time (e.g., approaching a calendar in the bedroom or preparing to exit through the front door), we proactively surface updated reminders of upcoming events, weather, or commute conditions. By aligning past interests with present routine, the system enables users to bypass repeated queries and supports smooth daily planning.

\para{Staying up-to-date in information-rich contexts (\hyperref[fig:usecase_and_applications]{A2})}
For activities involving regular information monitoring, such as research and entertainment, users often revisit similar referents and spaces with similar intentions (e.g., sitting at a workstation to look up new papers, or taking a break in a coffee corner while catching up on sports news). Recognizing these consistent contextual cues, the system retrieves previously asked queries (e.g., ``latest research papers'') and presents fresh responses situated on familiar objects in those spaces. This pairing of personal querying history with recurring usage locations turns routine contexts into a lightweight information provider.

\para{Recalling task-relevant instructions and reminders (\hyperref[fig:usecase_and_applications]{A3})}
Some tasks benefit from resurfacing specific information at the moment of re-engagement (e.g., technical maintenance steps or health-related reminders). When users return to the same tools, locations, or objects, the system proactively resurfaces prior instructions anchored to those referents. This helps reduce the cognitive overhead of re-remembering previously searched details and provides a just-in-time memory cue at the point of action.
\vspace{-0.5mm}

\section{Preliminary Evaluation}
\vspace{-0.5mm}
To assess the practicality and potential of proactive, context-aware AR assistance, we conducted a two-part preliminary evaluation: (1) a technical analysis of \sysmemento capabilities and performance, and (2) a ``take-home'' diary-based user study examining user acceptance, user experience, real-world usage patterns, and how the technology integrates into everyday routines. 
\vspace{-0.5mm}

\subsection{Technical Evaluation}
\label{subsec:technical_eval}

Identifying users' action contexts, and delivering timely and accurate situated AR content, are crucial for the practical adoption of proactive context-aware systems. In this section, we assess the context-aware capabilities of \sysmemento and report our results.

\para{Spatial activity context matching accuracy}
We evaluate the accuracy of the \hyperref[subsubsec:spatial_activity_recognizer]{Spatial Activity Recognizer} in identifying spatial and activity contexts from egocentric views using text-based scene representations. Each image is converted into a textual description with \gpt, and CLIP is used to generate embeddings directly from these textual summaries. To assess classification performance, we compute centroid embeddings for each spatial and activity class using a sampled dataset of 1,000 egocentric images drawn from Ego4D. For each test sample, its embedding is compared against class centroids using cosine similarity, retrieving the closest match as the predicted label.

The recognizer achieves an F1 score of 0.82 for Scene context classification and 0.82 for Activity context classification (see \cref{tab:spatial_activity_recog}). We find that this tendency continues until images contain motion blur or low lighting, where the accuracy drops to an F1 score of 0.57 and 0.30, respectively. These results indicate that text-based embeddings can provide semantic distinction to identify recurring spatial activities for proactive resurfacing, even without fine-tuning or visual modality fusion.

\para{Interaction referent recognition accuracy}
We found that the open-vocabulary object detector, YOLO-World-XL, occasionally struggles to robustly localize a referent that the user is interacting with. While it outperforms DetCLIP-T in speed (52 vs 2FPS) with similar accuracy (34.4 vs 35.4 AP), false positives led to faulty retrieval of \rsam and inconsistent situating of proactive AR visualizations, impacting the user experience.

To address this, we use \gemini as a fallback when detection confidence is low ($<0.5$). It is chosen for its spatial reasoning responsiveness ($\approx$$3.4s$) with competitive accuracy among other multimodal LLMs, at the time of the prototype design~\cite{patzold2025leveraging}.

\renewcommand{\arraystretch}{1.2}
\setlength{\tabcolsep}{1pt}
\begin{table}[!b]
\centering
\vspace{-4mm}
\caption{Performance comparison of CLIP variants--CNN-based (ResNet-101) and Transformer-based (ViT-B/32, ViT-L/14) architectures, for spatial Scene and Activity recognition tasks.}
\label{tab:spatial_activity_recog}
\begin{tabular}{|C{1.4cm}|p{2.85cm}|C{1.25cm}C{1.25cm}C{1.25cm}|}
\hline
\textbf{Task} & \textbf{CLIP Variants} & \textbf{Precision} & \textbf{Recall} & \textbf{F1-score}\\
\hline
 &
ViT-L-14 & 
\begin{tabular}{@{}ccc@{}}0.81\end{tabular}&
\begin{tabular}{@{}ccc@{}}0.88\end{tabular}&
\begin{tabular}{@{}ccc@{}}0.81\end{tabular}\\
Scene & ViT-B-32 & 
\begin{tabular}{@{}ccc@{}}0.80\end{tabular}&
\begin{tabular}{@{}ccc@{}}0.87\end{tabular}&
\begin{tabular}{@{}ccc@{}}0.80\end{tabular}\\
& ResNet-101 \textbf{(Ours)}& 
\begin{tabular}{@{}ccc@{}}0.82\end{tabular}&
\begin{tabular}{@{}ccc@{}}0.89\end{tabular}&
\begin{tabular}{@{}ccc@{}}\textbf{0.82}\end{tabular}\\
\hline
&
ViT-L-14 & 
\begin{tabular}{@{}ccc@{}}0.81\end{tabular}&
\begin{tabular}{@{}ccc@{}}0.85\end{tabular}&
\begin{tabular}{@{}ccc@{}}0.80\end{tabular}\\
Activity & ViT-B-32 & 
\begin{tabular}{@{}ccc@{}}0.75\end{tabular}&
\begin{tabular}{@{}ccc@{}}0.80\end{tabular}&
\begin{tabular}{@{}ccc@{}}0.73\end{tabular}\\
& ResNet-101 \textbf{(Ours)}&
\begin{tabular}{@{}ccc@{}}0.82\end{tabular}&
\begin{tabular}{@{}ccc@{}}0.87\end{tabular}&
\begin{tabular}{@{}ccc@{}}\textbf{0.82}\end{tabular}\\
\hline
\end{tabular}
\end{table}

\subsection{User Evaluation and Feedback}
\label{sec:user_evaluation}
\vspace{-1.25mm}
We collected data from 25 sessions conducted over a 2-4 day period, resulting in a total of $\approx$435 aggregated minutes, and 196 memories (the total number of query-generated memories across sessions; including user-dismissed memories) across all participants. Although each session lasted for 17.4 minutes on average, and was relatively short in duration, participants were exposed to 135 memories via proactive recalls (68.9\% of total memories).

\para{Participants and procedure}
We recruited 9 (P1-P9) self-volunteered participants (7 male, 2 female), aged 25-36 ($\mu$=29.1, $\sigma$=3.7), from Computer Science, Mathematics, and Physics backgrounds. Prior AR familiarity was broadly distributed on a 5-point scale (1:Not familiar, 5:Familiar), ($\mu$=2.8, $\sigma$=1.3), and gesture-based interaction experience showed a similar spread ($\mu$=3.1, $\sigma$=1.6). All participants reported regular daily use of personal computing devices. Participants' identities were anonymized, and each provided informed consent prior to participation. This study was conducted under the IRB approval of Stony Brook University ($1173920$). The participants were instructed to complete a brief pre-study questionnaire and a 30-minute introduction to \sysmemento. They were told to use \sysmemento naturally over 2-4 days in their daily indoor environments, with each usage counted as a ``session.'' After each session, participants submitted short diary entries describing context, perceived usefulness, and ease of integration, including a 5-point Likert scale and open-ended feedback. System logs of spatial context, interactions, and resurfaced memories were collected in parallel. After the multi-day use, participants completed a post-study survey comparing proactive and reactive support. Semi-structured interviews followed to explore how proactive cues affected routines and what limitations they encountered. To ensure consistency across participants, proactive resurfacing sensitivity was fixed at one resurfacing per 24 hours.

\para{Repeated daily usages}
Participants praised \sysmemento for its proactive task reminding ability--``\textit{I liked that I could place reminders around the house. Next to my shoes, I'll know I need to go out for a run}''(P3). However, one user reported referent misalignment: ``\textit{It mistook a square grocery bag to be a luggage and suggested me travel time}''(P3). A post-study interview revealed that the user moved close to the referent, enforcing our recognition model to infer with minimal visual context. In another case, a spatial activity was over-generalized: ``\textit{It said I'm gazing at an object. It's too generic}''(P2). We discuss candidate solutions in \cref{sec:limitation_discussion}. 

\vspace{-0.25mm}
\para{Acceptability in daily routine}
Users were favorable towards \sysmemento becoming part of their daily routine: ``\textit{Each time I use it, I get more comfortable with the interaction style. It feels like a solid foundation}''(P5). However, the requests to extend \sysmemento, to a lightweight glasses-based AR were suggested across participants (P1, P3, P4, P6, P8); ``\textit{It'll be nice to have a VR app integrated into lifestyle through glasses that pops up information quick}''(P6). A non-tech user said : ``\textit{It is cool, but I wear a huge glasses, so I can't wear it for long}''(P3). Furthermore, even with a multi-day AR experience, users expressed the need for a longer duration, to observe any changes in their routines: ``\textit{I cannot say it changed my routines yet. It clearly helped me, but I'd need to try it on for longer. I might feel differently if it is a glasses I can wear casually, like the Ray-ban}''(P1); ``\textit{Very promising system. Looking forward to seeing how it adapts with more use}''(P5). We extend our discussion to \cref{sec:limitation_discussion}. 
\vspace{-0.25mm}

\para{Proactive vs Reactive assistance}
Proactive context-aware AR in \sysmemento was evaluated positively: ``\textit{I check weather and traffic every morning, this gave me the answers before I checked}''(P7); ``\textit{It reduces a step in your daily patterns}''(P6); ``\textit{Great for reminders. I usually use Google Calendar, but many times my phone is on silent or put away, I'm unable to follow through. Proactivity definitely helps more and put bunch of To-Dos around my house}''(P3). Another participant compared \sysmemento to a traditional reactive assistant:  ``\textit{I use Google Nest Mini to voice query and get responses, but \sysmemento recall of previous queries is very useful}''(P1), and praised on its simplicity ``\textit{Biggest pros is that I don't need to think about setting up the conditions, just automatic and simple}''(P6).

\vspace{-1.5mm}
\section{Future Work and Discussion}
\label{sec:limitation_discussion}
\vspace{-1mm}
\para{Towards everyday AR form factors}
Participants recognized the usefulness of proactive resurfacing but highlighted the mismatch between the system's value and its current hardware. As one noted, ``\textit{If it was like glasses, I could really see the benefits of AR outdoors. It's awkward to go out with this}'' (P4). Everyday deployment likely depends on lightweight glasses-class devices such as Meta's Ray Ban glasses or XReal~\cite{xreal}, where proactive information blending can occur with less social discomfort. Since the system already offloads computation to remote servers, cross-platform deployment is feasible. Future work will explore non-verbal triggers such as audio, gaze, and micro-gestures, building on research into subtle input for public AR use~\cite{lu2021evaluating, lu2023wild}, enabling proactive retrieval even when speech would be socially intrusive.
\vspace{-0.25mm}

\para{Mixed-initiative context control and activity granularity}
Accurate resurfacing depends on inferring when a resurfaced memory is contextually appropriate. AR content that is mis-triggered may negate the usefulness and utility of proactive visualization. While participants generally appreciated autonomous triggers, they wanted occasional corrective control: ``\textit{I sometimes wanted more control}''(P5). This suggests that proactive AR should adopt mixed-initiative mechanisms that remain simple, yet adjustable. Prior work on user-assistant collaboration and conditional triggers, such as IFTTT (If-This-Then-That)~\cite{ifttt, shakeri2023user}-like structures, may provide contexts for lightweight refinements. Improving recognition accuracy may also benefit from hierarchical activity representations as suggested by ongoing efforts in structured activity inference~\cite{edge2024local, userllm, xie2024embodied, yang2025egolife}, mitigating over-generalization and helping maintain alignment between resurfacing timing and user intent.
\vspace{-0.25mm}

\para{Continuity of context across realities}
Participants responded positively when resurfaced information appeared, ``\textit{where it felt natural},'' especially when memories were tied to objects already associated with an activity. While the current panel-based overlays provide invariance to layout changes, they lack deeper geometric coupling with the environment. Closer integration between AR content and spatial referents could strengthen relevance and improve interpretability and explainability, building on design considerations for spatially anchored AR interfaces~\cite{lee2023design}. Future work should explore referent-aware embeddings that balances spatial robustness with meaningful anchoring.

\para{Need for longitudinal routine pattern evaluation}
Our diary-based deployment captured short-term routine patterns but could not reveal how resurfacing should adapt as habits evolve. Participants occasionally treated the system as a temporary ``study task,'' leading to usage that did not fully reflect everyday rhythms. Longer-term field studies would allow users to internalize proactive AR as part of daily life, similar to long-term wearable research~\cite{bressa2022data}. Understanding when proactive reminders should scale back, pause, or revive requires extended observation of routine drift and habit change over weeks or months, rather than days.
\vspace{-0.25mm}

\para{Privacy boundaries and memory retention}
Even though the system did not store raw imagery, participants raised concerns about what the system ``remembers'' and whether their home environment might be inferred from resurfaced cues: ``\textit{Do you save the captures? I don't want my room to be seen}''(P4). This aligns with longstanding concerns about visual privacy in lifelogging systems~\cite{azuma2019road} and underscores the need for user control over memory retention and deletion. Prior work on selective access control and user agency~\cite{kim2021design, erebus, rajaram2023reframe, rajaram2025exploring} suggests that proactive systems require transparency mechanisms and forgetting controls that allow resurfacing without unintended exposure of personal spaces or objects.

\para{Systematic evaluation of proactive benefits}
This work explored proactive resurfacing as a design space, but the cognitive and performance effects compared to reactive assistants remain under-examined. In our future work, we plan to investigate whether resurfaced memories reduce recall burden or task-switching overhead, and examine how multimodal cues such as visual, audio, and gaze combinations influence attention and action~\cite{mayer2002multimedia, lee2025walkie, cai2025aiget}. Understanding when proactivity assists, distracts, or overwhelms will refine how proactive AR integrates into everyday practice.
\vspace{-1mm}

\section{Conclusion}
\vspace{-0.25mm}
To explore the usability of proactive, context-aware assistance on a wearable AR device, we designed a foundational system, \sysmemento, that captures and stores users' verbal interactions as Referent-anchored Spatiotemporal Activity Memories, and proactively resurfaces them based on recurring daily contexts. We introduced a hybrid data retrieval structure combining spatial indexing and semantic vector search, and implemented a simplified context-aware system by integrating multimodal sensing, LLM-based reasoning, and open-vocabulary referent detection. Through real-world use cases, we investigated the potential for everyday use of a proactive AR assistant, and discuss takeaways and future work.

\acknowledgments
{
The initial sketch of \cref{fig:memento_design_concept} was edited with the help of the Images model of ChatGPT; This research was supported in part by NSF award IIS2529207, IIS2441310 and ONR award N000142312124.
}

\bibliographystyle{abbrv-doi-hyperref-narrow}

\bibliography{references}
\end{document}